\documentclass[namedreferences]{solarphysics}
\usepackage[optionalrh,solaromanenum]{spr-sola-addons} 
\usepackage{graphicx}                    
\usepackage{epsfig}
\usepackage{epstopdf}
\usepackage{color}                       
\usepackage{url}                         

\usepackage{natbib}

\begin{document}

\def\aj{AJ}%
\def\actaa{Acta Astron.}%
\def\araa{ARA\&A}%
\def\apj{Astrophys. J.}%
\def\apjl{Astrophys. J.}%
\def\apjs{ApJS}%
\def\ao{Appl.~Opt.}%
\def\apss{Ap\&SS}%
\def\aap{Astron. Astrophys.}%
\def\aapr{Astron. Astrophys. Rev.}%
\def\aaps{A\&AS}%
\def\azh{AZh}%
\def\baas{BAAS}%
\def\bac{Bull. astr. Inst. Czechosl.}%
\def\caa{Chinese Astron. Astrophys.}%
\def\cjaa{Chinese J. Astron. Astrophys.}%
\def\icarus{Icarus}%
\def\jcap{J. Cosmology Astropart. Phys.}%
\def\jgr{J. Geophys. Res. (Space Physics)}
\def\jrasc{JRASC}%
\def\mnras{MNRAS}%
\def\memras{MmRAS}%
\def\na{New A}%
\def\nar{New A Rev.}%
\def\pasa{PASA}%
\def\pra{Phys.~Rev.~A}%
\def\prb{Phys.~Rev.~B}%
\def\prc{Phys.~Rev.~C}%
\def\prd{Phys.~Rev.~D}%
\def\pre{Phys.~Rev.~E}%
\def\prl{Phys.~Rev.~Lett.}%
\def\pasp{PASP}%
\def\pasj{Publications of the Astronomical Society of Japan}%
\def\qjras{QJRAS}%
\def\rmxaa{Rev. Mexicana Astron. Astrofis.}%
\def\skytel{S\&T}%
\def\solphys{Solar~Phys.}%
\def\sovast{Soviet~Ast.}%
\def\ssr{Space~Sci.~Rev.}%
\def\zap{ZAp}%
\def\nat{Nature}%
\def\iaucirc{IAU~Circ.}%
\def\aplett{Astrophys.~Lett.}%
\def\apspr{Astrophys.~Space~Phys.~Res.}%
\def\bain{Bull.~Astron.~Inst.~Netherlands}%
\def\fcp{Fund.~Cosmic~Phys.}%
\def\gca{Geochim.~Cosmochim.~Acta}%
\def\grl{Geophys.~Res.~Lett.}%
\def\jcp{J.~Chem.~Phys.}%
\def\jgr{J.~Geophys.~Res.}%
\def\jqsrt{J.~Quant.~Spec.~Radiat.~Transf.}%
\def\memsai{Mem.~Soc.~Astron.~Italiana}%
\def\nphysa{Nucl.~Phys.~A}%
\def\physrep{Phys.~Rep.}%
\def\physscr{Phys.~Scr}%
\def\planss{Planet.~Space~Sci.}%
\def\procspie{Proc.~SPIE}%
\let\astap=\aap
\let\apjlett=\apjl
\let\apjsupp=\apjs
\let\applopt=\ao

\begin{article}

\begin{opening}

\title{Comparative Study of MHD Modeling of the Background Solar Wind}

%
\author{C.~\surname{Gressl}$^{1}$\sep
        A.\,M.~\surname{Veronig}$^{1}$\sep
        M.~\surname{Temmer}$^{1}$\sep
        D.~\surname{Odstr\v{c}il}$^{2}$\sep
        J.\,A.~\surname{Linker}$^{3}$\sep
        Z.~\surname{Miki\'{c}}$^{3}$\sep
        P.~\surname{Riley}$^{3}$          
               }

%
\runningauthor{C.\ Gressl \textit{et al.}}
\runningtitle{Comparative Study of MHD Modeling of the Background Solar Wind}

%
  \institute{$^{1}$ Kanzelh\"ohe Observatory/IGAM, Institute of Physics, University of Graz, Austria, email: \url{corinna.gressl@uni-graz.at} \\ 
             $^{2}$ George Mason University, Fairfax, VA, USA \\                  
             $^{3}$ Predictive Science, Inc, San Diego, CA, USA \\
             }

\begin{abstract}
Knowledge about the background solar wind plays a crucial role in the framework of space weather forecasting. \textit{In-situ} measurements of the background solar wind are only available for a few points in the heliosphere where spacecraft are located, therefore we have to rely on heliospheric models to derive the distribution of solar wind parameters in interplanetary space. We test the performance of different solar wind models, namely Magnetohydrodynamic Algorithm outside a Sphere/ENLIL (MAS/ENLIL), Wang--Sheeley--Arge/ENLIL (WSA/ENLIL), and MAS/MAS, by comparing model results with \textit{in-situ} measurements from spacecraft located at 1~AU distance to the Sun (ACE, \textit{Wind}). To exclude the influence of interplanetary coronal mass ejections (ICMEs), we chose the year 2007 as a time period with low solar activity for our comparison. We found that the general structure of the background solar wind is well reproduced by all models. The best model results were obtained for the parameter solar wind speed. However, the predicted arrival times of high-speed solar wind streams have typical uncertainties of the order of about one~day. Comparison of model runs with synoptic magnetic maps from different observatories revealed that the choice of the synoptic map significantly affects the model performance.
\end{abstract}

%

\end{opening}


\section{Introduction}
\label{sec:introduction}

High-speed solar wind streams (HSSs) play an important role in space weather at Earth. Recurrent geomagnetic storms are strongly associated with HSSs (\citeauthor{crooker1994}, \citeyear{crooker1994}, and references therein), as are high-energy electrons in the Earth's magnetosphere \citep{baker1997}. HSSs emanate from coronal holes on the Sun and form corotating interaction regions (CIRs) with enhanced densities and magnetic field strengths \citep{tsurutani2006}.
The most severe geomagnetic storms are caused by coronal mass ejections (CMEs) that occur frequently during times of high solar activity.
The interplanetary propagation of CMEs is strongly influenced by the interaction with the ambient solar wind flow, which can either accelerate or decelerate the CME (\textit{e.g.} \citeauthor{gopalswamy2000}, \citeyear{gopalswamy2000}; \citeauthor{temmer2011}, \citeyear{temmer2011}). To estimate the potential geoeffectiveness of CMEs as well as their arrival time at the Earth, information on their speed and propagation direction as well as on the ambient solar wind flow is needed.
Therefore, knowledge of the background solar wind is a key ingredient in the framework of space weather forecasting.

Measurements of the background solar wind conditions are only available for a few points in the heliosphere, \textit{e.g.} measurements from ACE and \textit{Wind}, which are located at Lagrangian point L$_1$ at $\approx$1~AU in the Sun--Earth line, about one hour upstream of the Earth.
To gain information on the background solar wind distribution between the Earth and the Sun, we need to rely on numerical MHD modeling or on empirical relationships (\textit{e.g.} \citeauthor{vrsnak2007}, \citeyear{vrsnak2007}, \citeauthor{rotter2012}, \citeyear{rotter2012}).
Numerical solar wind models typically simulate the heliosphere in the distance range from about 20~R$_{\odot}$ to 2~AU and provide simulations of solar wind parameters such as speed, density, temperature, and magnetic field strength.

The solar wind models are constantly improved and updated by the code developers. Validation and testing of the model performance is an important step in further improving the models in order to ensure an accurate description of our space environment. Solar wind models can be tested by comparing the model output at the location of a spacecraft to actual \textit{in-situ} measurements from the spacecraft.

\cite{owens2008} tested the Wang--Sheeley--Arge/ENLIL (WSA/ENLIL) and Magnetohydrodynamic Algorithm outside a Sphere/ENLIL (MAS/ENLIL) models over the time period 1995\,--\,2002, a time period that ranges from low (1995) to high solar activity (maximum of Cycle 23 in 2002). These authors found a good model performance for large-scale structures and a systematic time offset for these models of about two days.
In another recent study, \cite{lee2009} compared the solar wind models MAS/ENLIL and WSA/ENLIL with \textit{in-situ} measurements from ACE for Carrington rotations (CRs) 1999\,--\,2038 (24 January 2003\,--\,18 January 2006). To this end, the data were extracted from the simulation output in the ecliptic plane at a fixed distance to the Sun (1~AU). These authors found a general good agreement between the solar wind models and the \textit{in-situ} measurements for large-scale structures and for time scales of several days.
\cite{jian2011} compared MAS/ENLIL and WSA/ENLIL model results with observations at ACE and Ulysses for CRs 2016\,--\,2018 when ACE and Ulysses were in latitudinal alignment. The alignment made it possible to compare the model results for the same time and latitude but for different distances (1~AU and 5.4~AU) to the Sun. These authors found that for the two-stream interaction regions during CRs 2016\,--\,2018 the models are able to simulate field polarities, sector boundaries, and the occurrence and features of stream-interaction regions. However, the simulated arrival times of these structures were off by about two days.

In this article, we present an evaluation of the solar wind models MAS/MAS, MAS/ENLIL, and WSA/ENLIL during a time period when solar activity was exremely low (year~2007). For the prediction of transit times of CMEs from the Sun and their impact speed at Earth, background solar wind models that produce accurate results on time scales less than one day are needed. Therefore, we compared the model results with \textit{in-situ} measurements at the exact location of the spacecraft on timescales of several hours to days.

\section{Solar Wind Models}

Today, the approach for heliospheric modeling is the use of coupled models, where the simulations of the corona are separated from the simulations of the heliosphere. The coronal part of the model uses magnetic synoptic maps as input parameter and extrapolates it to a source surface at typically 2.5~R$_{\odot}$. The synoptic maps are built up from line-of-sight measurements of the Sun's photospheric magnetic field over the course of a solar rotation. From the source surface at 2.5~R$_{\odot}$ out to about 20\,--\,30~R$_{\odot}$ the coronal part of the model simulates the conditions in the corona \citep{riley2001}. The outer boundary conditions derived by the coronal model is then used as the inner boundary condition for the heliospheric component of the model, that simulates the solar wind properties out to 1~AU and beyond.

Coronal solutions in MAS (1\,--\,30~R$_{\odot}$) solve the three dimensional (3D) MHD equations in spherical coordinates on nonuniform grids \citep{linker1999, mikic1999}. Boundary conditions for the polytropic version of MAS include the radial magnetic field at the photosphere as supplied from synoptic magnetic maps (available from a number of observatories) and the plasma density and temperature at the base of the corona (see \citeauthor{linker1999}, \citeyear{linker1999} for more details about the solution procedure). Solutions with a more advanced energy equation \citep{lionello2009} are also available at the PSI web site for rotations after the launch of SDO (\url{www.predsci.com/hmi/}) but were not used in this study. The radial magnetic field from the coronal solution is used as the boundary condition for the heliospheric solution (inner boundary at 30~R$_{\odot}$); the heliospheric solution can be computed with either a heliospheric version of the MAS code (MAS/MAS) or the ENLIL model (MAS/ENLIL, also called CORHEL). In either case, an empirical prescription is used for setting the plasma boundary conditions in the heliospheric model. The radial velocity of the solar wind at 30 R$_{\odot}$) is calculated  assuming that high-speed flows originate from coronal holes, and low-speed flows from the boundaries of coronal holes. The proton density is derived from momentum-flux balance, and plasma temperature is calculated under the assumption of thermal-pressure balance at the inner radial boundary. The boundary conditions are assumed to be steady in the rotating frame of the Sun. The MAS model neglects possible effects of pick-up ions, and is therefore limited to a modeling range $<$\,5~AU. A more detailed description of MAS heliospheric solutions is given by \cite{riley2001}.


ENLIL is a time-dependent 3D MHD heliospheric model to simulate the structure and evolution of the solar wind. ENLIL is suited for simulations of the inner and middle heliosphere and can be used to model the background solar wind as well as transient disturbances in the heliosphere such as CMEs \citep{odstrcil2002, odstrcil2003}.
ENLIL uses an ideal-fluid approximation and solves for the equations of ideal magnetohydrodynamics using a Total Variation Diminishing Lax Friedrich Scheme algorithm (\url{ccmc.gsfc.nasa.gov/models/modelinfo.php?model=ENLIL}). It assumes equal densities and temperatures for electrons and protons and neglects microscopic processes. 
The inner boundary conditions can either be retrieved from MAS (MAS/ENLIL) or WSA (WSA/ENLIL).
Starting from the inner boundary, the ENLIL code then simulates the propagation of solar wind structures outward into the heliosphere. For simulations of the near-Earth environment, it is recommended to set the outer boundary of the ENLIL code to 2~AU. To include spacecraft and planets in the outer heliosphere, the ENLIL code also offers the possibility of being run out to 10~AU.


The WSA model \citep{wang-sheeley1990, arge-pizzo2000} is a semi-empirical model that relates the expansion factor of magnetic flux tubes in the corona to solar wind speeds at 1~AU. It is based on the findings of \cite{levine1977} that high-speed solar wind streams are correlated with low magnetic flux tube expansions between the photosphere and the corona.
For the computation of the coronal magnetic field up to 2.5~R$_{\odot}$ the potential-field source-surface model
 \citep[PFSS:][]{levine1977} is used, beyond that the Schatten current sheet model \citep{schatten1969}. 
The plasma density and temperature are calculated from the momentum flux and pressure balance at the inner boundary.

ENLIL and MAS produce stationary solutions for the background solar wind. The MAS model consists of both a coronal and heliospheric component, while ENLIL is a purely heliospheric model that needs input from the coronal component of either the MAS or the WSA model to obtain the inner boundary conditions. ENLIL model runs can be requested on the homepage of NASA's Community Coordinated Modeling Center (CCMC) and are run on demand for the user. Heliospheric models are also provided by the Space Weather Modeling Framework (SWMF) of the University of Michigan. The SWMF computes global  MHD solutions that are driven by observed photospheric magnetic fields \citep{toth2005}.

\section{Data and Methods}

The models tested in this study simulate the background solar wind, \textit{i.e.} the solar wind conditions during quiet times of the solar-activity cycle without consideration of disturbances from transient solar events such as CMEs. To minimize the effect of transient events on the model performance and the comparison between model and data, we have selected the year 2007 as a time period with low solar activity during the declining phase of Cycle~23.
For the year 2007, only two interplanetary coronal mass ejections (ICMEs) were identified in \textit{in-situ} plasma and field measurements at 1 AU in the Sun--Earth line as, \textit{e.g.}, opposed to 51 ICMEs during the year 2001, when Cycle~23 had its maximum (as inferred from the ICME list of \citeauthor{richardson2010}, \citeyear{richardson2010}).

The MAS/MAS model runs were performed by Predictive Science Inc.\ and the model results were provided in the form of \textsf{hdf}-files on the Predictive Science homepage: \url{www.predsci.com/stereo/dataAccess.php}. For the ENLIL model runs, we used the online-request service on the webpage of CCMC: \url{ccmc.gsfc.nasa.gov/index.php}. The model runs were carried out by CCMC and the data of the model results were provided online on the web page in form of \textsf{netCDF} files.
We also used the option to run the WSA/ENLIL model with different sources for the synoptic magetic maps (from the \textit{National Solar Observatory} (NSO), the \textit{Mount Wilson Observatory} (MWO), and from the \textit{Global Oscillation Network} (GONG)). For the comparison with the observational data, we used data from \textsf{hdf}-files that were provided on special request and contained the full simulation data. 
The model versions used at CCMC at the time that we requested the model runs, were \textsf{ENLIL version~2.6} and \textsf{CORHEL-4.2.r45} for the MAS/ENLIL model runs and \textsf{ENLIL~2.7} and \textsf{WSA 2.2} for the WSA/ENLIL model runs. For the MAS/MAS model runs that were performed by Predictive Science, \textsf{CORHEL version 4.7.0} was used. The MAS/ENLIL solutions from CCMC come from an earlier version of CORHEL and the procedures for processing the synoptic maps differ from the MAS/MAS solutions from Predictive Science.

		\begin{figure}
			\centering
			\includegraphics[width=1\textwidth]{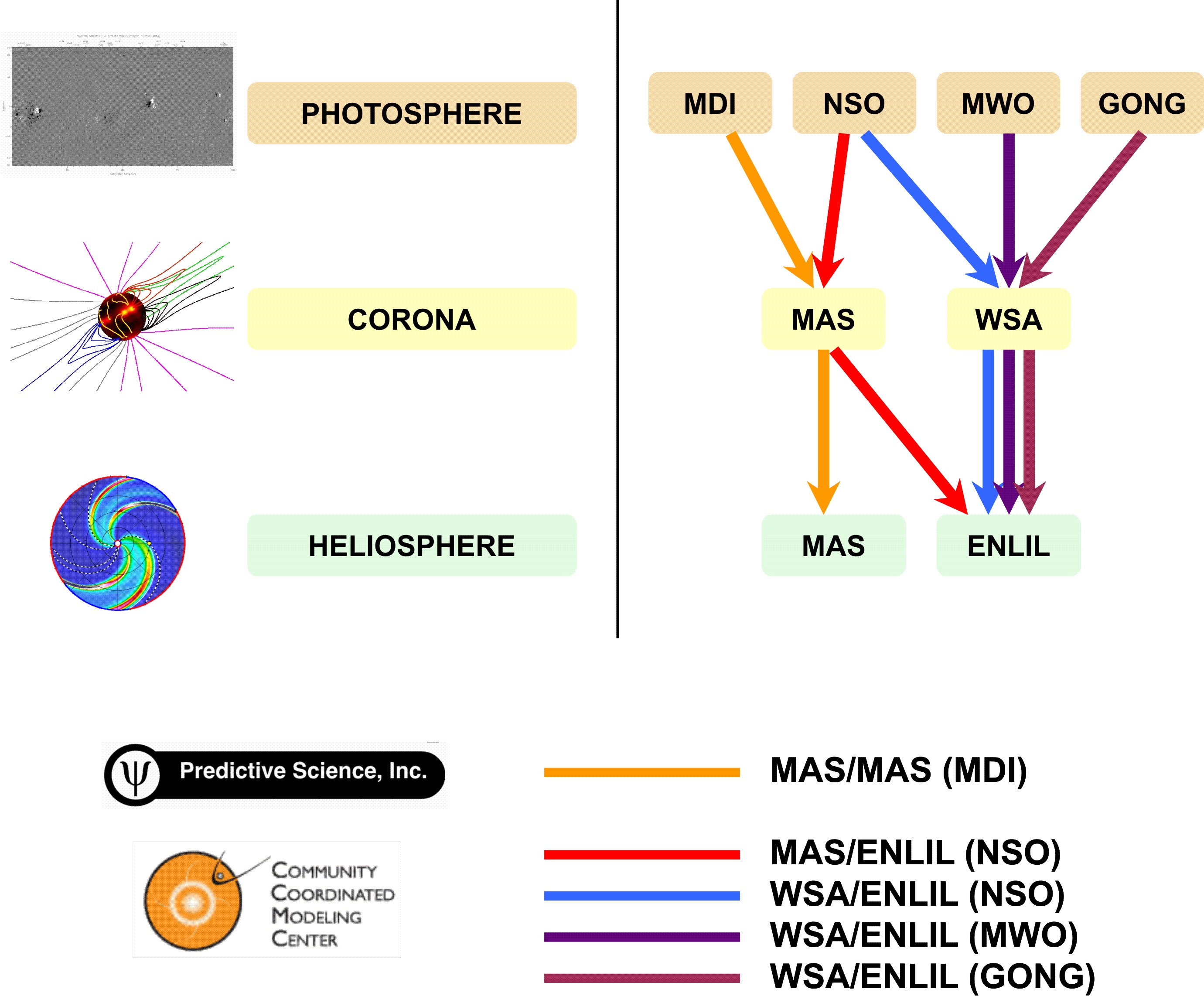}
			\caption{Different combinations of coronal and heliospheric models (MAS, WSA, ENLIL) as well as sources for the synoptic map (MDI, NSO, MWO, GONG) used in this study.}
			\label{fig:flow_diagramm_models}
		\end{figure}

Altogether, for each CR five different model runs were performed. We tested the following combinations of coronal/heliospheric models against \textit{in-situ} observations in the Sun--Earth line at 1~AU (see also the illustration in Figure~\ref{fig:flow_diagramm_models}):
\begin{itemize}
	\item MAS/MAS with synoptic maps from MDI,
	\item MAS/ENLIL with synoptic maps from NSO,
	\item WSA/ENLIL with synoptic maps from NSO,
	\item WSA/ENLIL with synoptic maps from MWO, and
	\item WSA/ENLIL with synoptic maps from GONG.
\end{itemize}

\begin{figure}[tb]
		\includegraphics[width=1\textwidth]{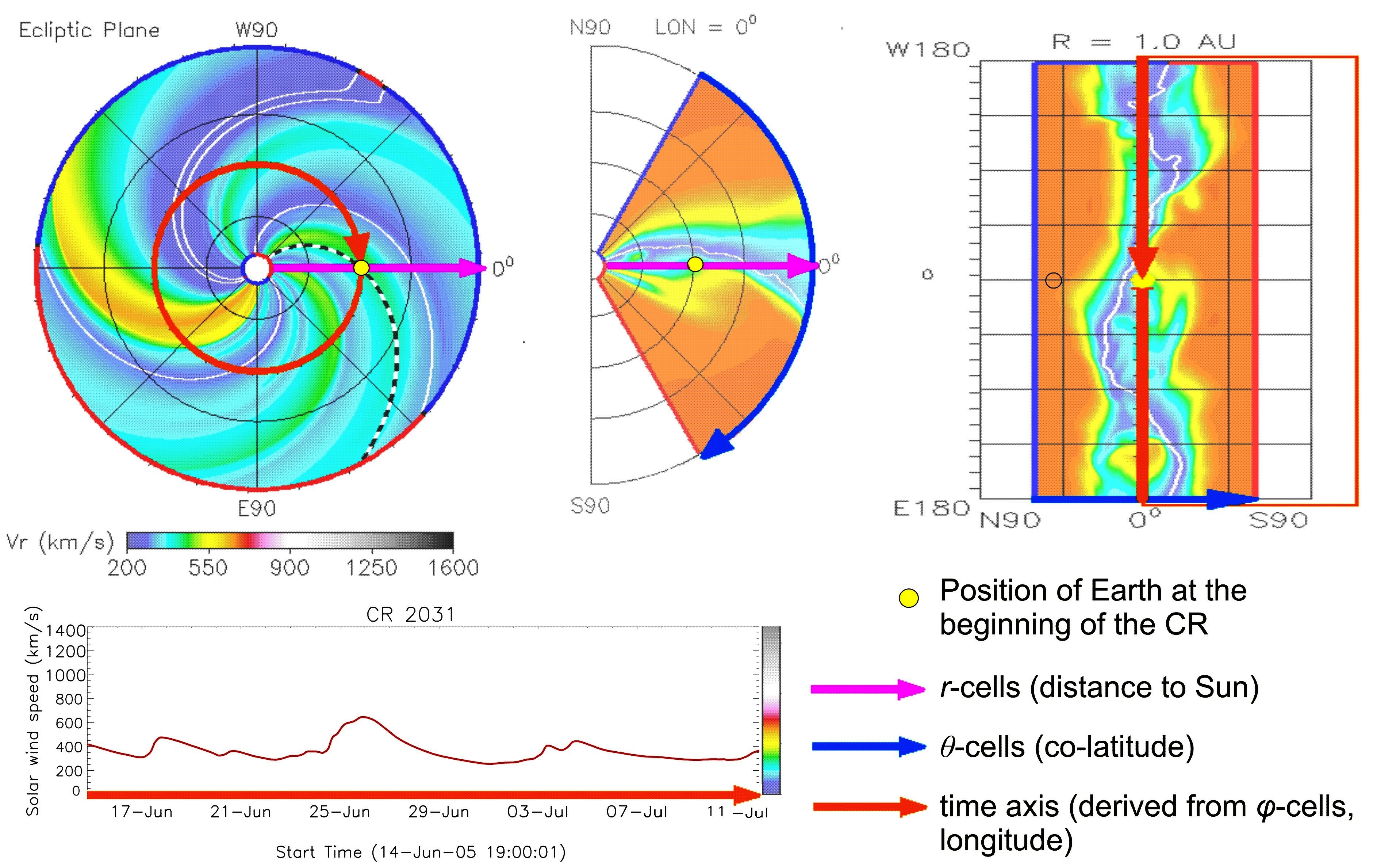}
			\caption{Quick-look graphics for the model runs showing the location of the $r$, $\theta$, and $\phi$ cells in the ENLIL simulation grid. Note that the longitudinal cells run in counter-clockwise direction and have the opposite direction as the time axis.
							}
			\label{fig:cells}
	\end{figure}

For the ENLIL simulations, a grid resolution of 1024~$\times$~120~$\times$~360 ($r$~$\times$~$\theta$~$\times$~$\phi$; see Figure~\ref{fig:cells} for an illustration of the simulation cells) was chosen. The $r$-cells cover 1024 pixels in radial distance~[r] from the inner boundary (located at 30.3~R$_{\rm{\odot}}$ for MAS/ENLIL, and at 21.7~R$_{\rm{\odot}}$ for WSA/ENLIL) to the outer boundary at 2~AU.
The 120 latitudinal cells [$\theta$] cover $\pm$60$^{\circ}$ in latitude. The 360 longitudinal cells [$\phi$] range from 0$^{\circ}$ to 360$^{\circ}$. The 360 longitudinal cells correspond to a time span of one Carrington rotation (27.27~days) and were converted to a time axis with a resolution of $\approx$1.8~hours, with the first cell at 0$^{\circ}$ corresponding to the start date of the CR. For the MAS/MAS model, the \textsf{hdf}-files have a size of 140~$\times$~111~$\times$~129 ($r$~$\times$ $\theta$~$\times$ $\phi$) and the 129 longitudinal cells were converted to a time axis with a resolution of $\approx$five hours. The different spatial resolution of the models result in a temporal resolution of 1.8~hours (MAS/ENLIL and WSA/ENLIL) and $\approx$five~hours (MAS/MAS). The spatial resolution of the models does not allow us to simulate small-scale structures in the solar wind. However, for this study, we focus on large-scale structures, like high-speed solar wind streams, which are quasi-steady and thus more amenable to equilibrium modeling. The spatial and temporal resolution of the models is sufficient for the purpose of modeling large-scale structures in the background solar wind. For retrieving the data of the model results, we extracted the solar wind parameters from the \textsf{hdf}-files for the actual position of the ACE and \textit{Wind} spacecraft [L$_{\rm{1}}$] with a procedure that compares the spacecraft's coordinates with the location of the simulation cells and determines the best matching cell in the simulation grid, taking into account also the radial and latitudinal changes of the spacecraft's orbit during the course of one CR.

For the observational data, against which the models were tested, we used hourly averaged \textit{in-situ} data from \textit{Wind} \citep{gloeckler1995} and the \textit{Advanced Composition Explorer} \citep[ACE:][]{stone1998}. For the speed, density, and temperature parameters we used data from the \textit{Solar Wind Experiment} (SWE: \citeauthor{ogilvie1995}, \citeyear{ogilvie1995}) onboard the \textit{Wind} spacecraft.
Data from the \textit{Solar Wind Electron Proton Alpha Monitor} (SWEPAM: \citeauthor{mccomas1998}, \citeyear{mccomas1998}) onboard ACE were used to fill data gaps in the SWE speed, density, and temperature data. For the parameters radial and total magnetic field strength we used data from the magnetic field experiment (MAG: \citeauthor{smith1998}, \citeyear{smith1998}) onboard ACE.

We calculated correlation coefficients to quantify the agreement between modeled and \textit{in-situ} measured solar wind parameters. In order to assess the significance of the correlations and to obtain a robust estimate of the correlation coefficients, we applied bootstrapping \citep{wall2003}. The bootstrap method works in the following way: Out of the sample of $N$ events, we draw repeatedly $N$ events at random, and compute the correlation coefficient for each of these realizations. This procedure is repeated 1000 times, and the average value of the distribution of the correlation coefficients is calculated.

\section{Results}

We tested the performance of different solar wind models against \textit{in-situ} measurements with respect to their use in space weather forecasts. Therefore, we were mainly interested in two aspects: First, the general agreement between modeled and measured solar wind parameters, \textit{i.e.} how well do the models reproduce characteristic solar wind structures such as HSSs or CIRs. Second, we were interested in how well the models forecast the arrival times and amplitudes of solar wind structures at the Earth with particular focus on the parameter solar wind speed. We evaluated the model results both for single CR model runs and for the whole time period of the year 2007 (CR~2052\,--\,2065). To quantify the agreement between model results and observations, we calculated the Pearson-Bravais correlation coefficient, which quantifies the linear relationship between two parameters.

\subsection{Correlation Analysis}

\begin{figure}
			\centering
			\includegraphics[width=1\textwidth]{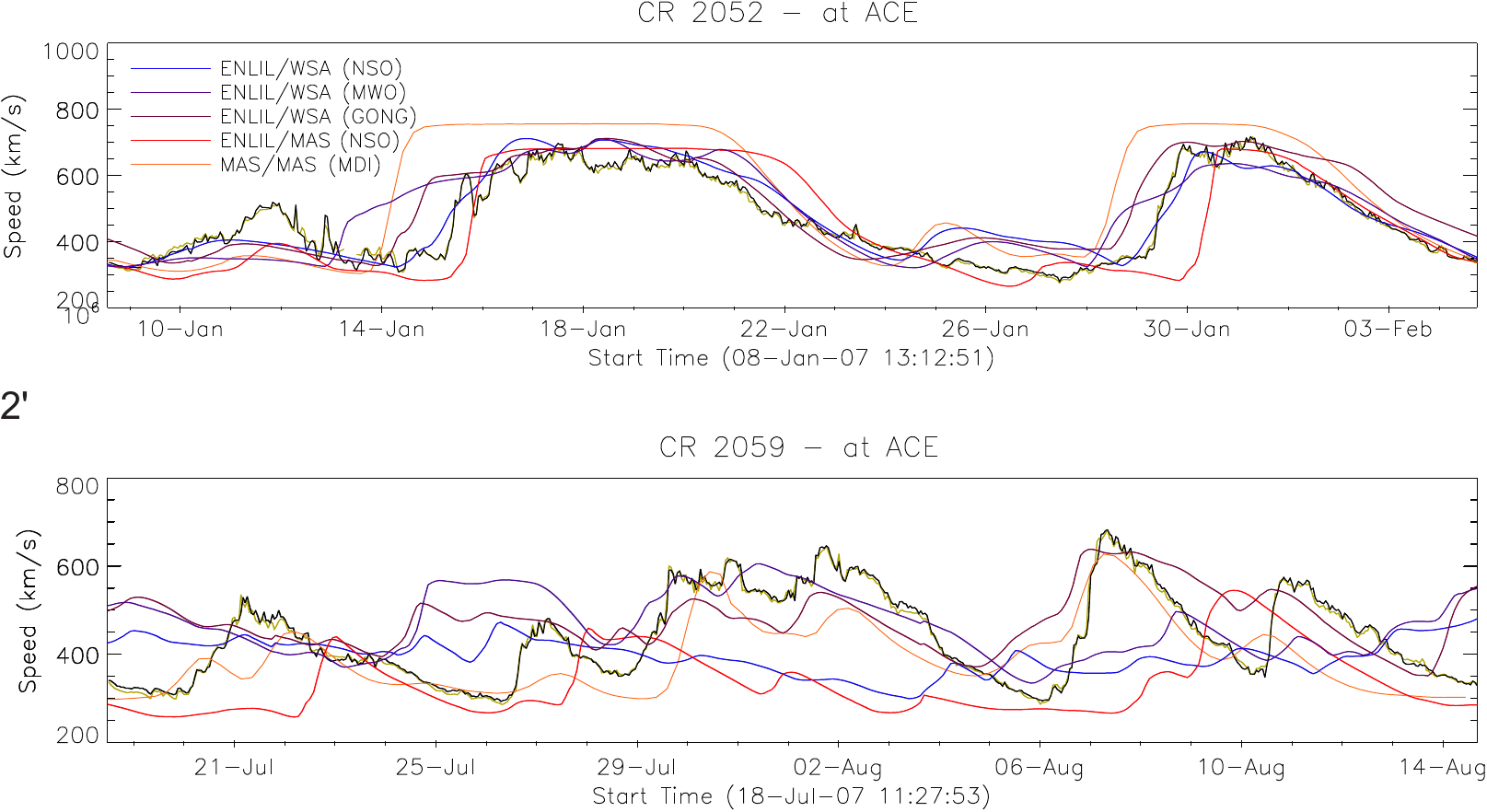}
			\caption{Sample model results and \textit{in-situ} observations of the solar wind speed for CR~2052 (top) and CR~2059 (bottom). \textit{In-situ} measurements are plotted by black lines, MAS/ENLIL model results in red, WSA/ENLIL (NSO) model results in blue, WSA/ENLIL (MWO) model results in purple, WSA/ENLIL (GONG) model results in dark red, and MAS/MAS (MDI) model results in orange.}
			\label{fig:speed_2005_2007}
		\end{figure}

Results of the model runs together with \textit{in-situ} data for solar wind speed are shown in Figure~\ref{fig:speed_2005_2007} as an example for two Carrington rotations. For CR~2052 (top panel) the models reveal a good agreement with the observations, with high correlation coefficients (cc\,$\approx$0.8). The modeled maximum speeds of the two HSSs as well as the speeds of the intermediate slow solar wind are in basic agreement with the observations. However, the predicted arrival times for the HSSs differ from model to model. For CR~2059 (bottom panel) the model results show very low agreement with the \textit{in-situ} measurements: the maximum value of solar wind speed enhancements are not well reproduced. Also the shape of the structures and their arrival times are not well captured by the models.

\begin{figure}
			\centering
			\includegraphics[width=1\textwidth]{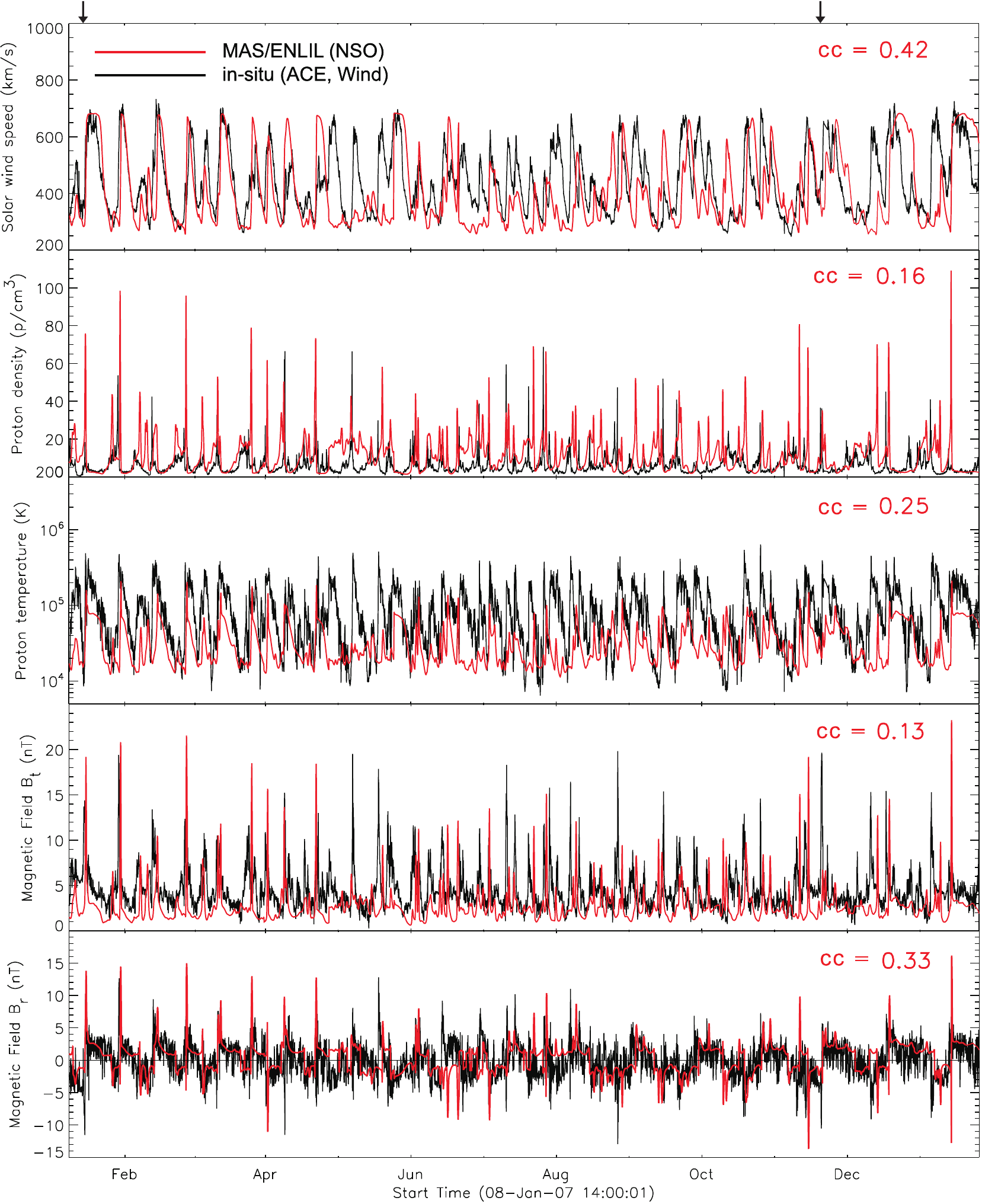}
			\caption{Modeled solar wind parameters from MAS/ENLIL (NSO, red) and the \textit{in-situ} observations (black) for CRs~2052\,--\,2065. From top to bottom: solar wind speed, proton density, proton temperature, total magnetic field strength, and radial magnetic field strength. The arrows on top indicate the start time of ICMEs identified at 1~AU (from \citeauthor{richardson2010}, \citeyear{richardson2010}).}
			\label{fig:mas_all_parameters}
		\end{figure}

\begin{figure}
			\centering
			\includegraphics[width=1.0\textwidth]{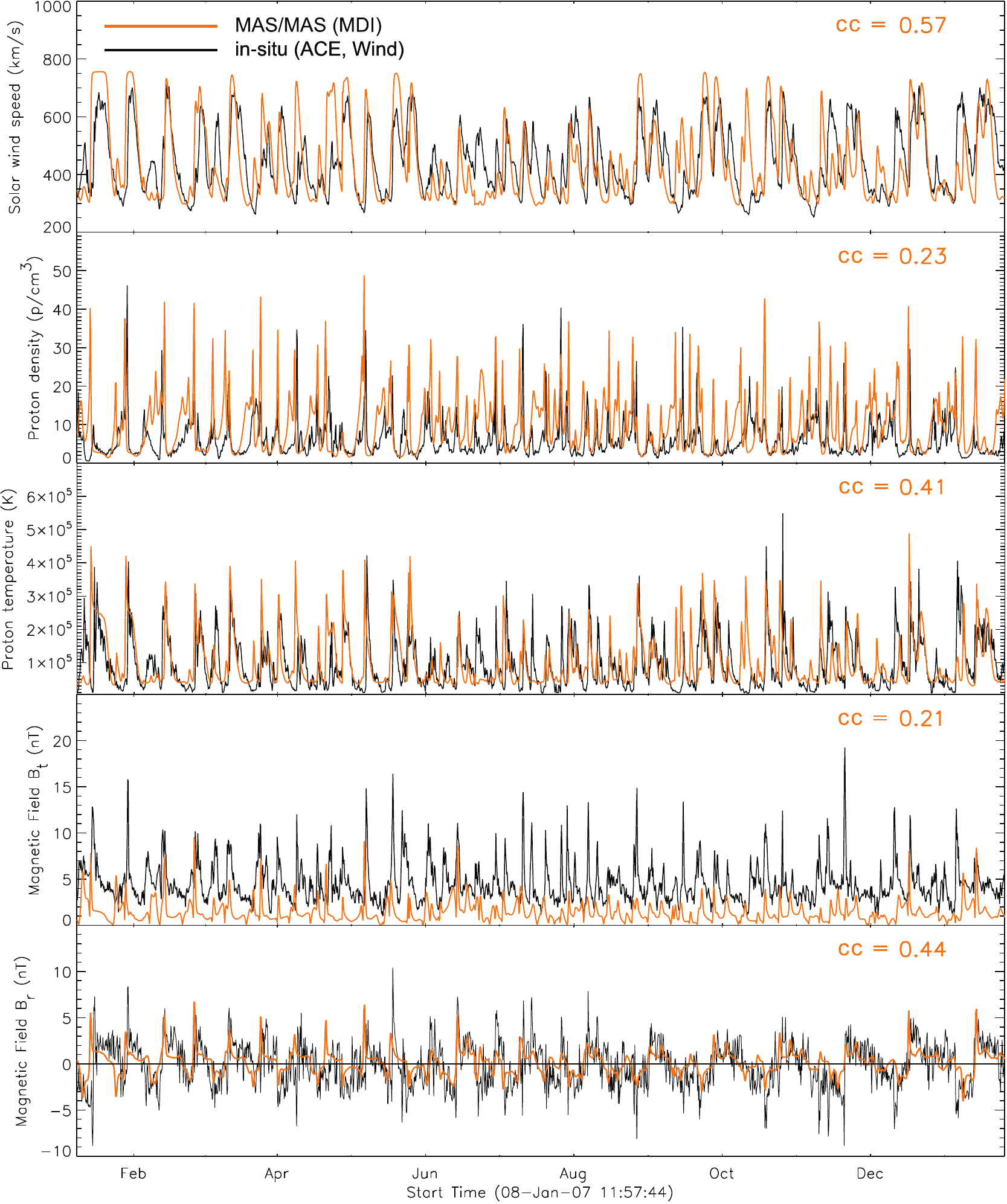}
			\caption{Modeled solar wind parameters from MAS/MAS (MDI, orange) and the \textit{in-situ} observations (black) for CRs~2052\,--\,2065. From top to bottom: solar wind speed, proton density, proton temperature, total magnetic field strength, and radial magnetic field strength.}
			\label{fig:pred_all_parameters}
		\end{figure}

Our comparison of the model runs throughout the year 2007 are presented in Figures~\ref{fig:mas_all_parameters}\,--\,\ref{fig:pred_all_parameters}. The figures show the model results as well as the \textit{in-situ} measurements from \textit{Wind} and ACE for all parameters under study: solar wind speed, proton density, temperature, total magnetic field strength, and radial magnetic field strength.
For all different combinations of coronal/heliospheric models, the best model performance was obtained for the solar wind speed parameter with a correlation coefficient of cc~=~0.42 for MAS/ENLIL, cc~=~0.53 for WSA/ENLIL, and cc~=~0.57 for MAS/MAS.
Good results were also obtained for simulating the sector structure of the heliosphere, \textit{i.e.} the radial magnetic field strength, with correlation coefficients between 0.3 and 0.5. However, the modeled radial magnetic field strength is in general smaller than what can be found in the measurements.
For the parameter proton temperature we find correlation coefficients between 0.25 and 0.41. For MAS/MAS the modeled peak values for the temperature are comparable to the \textit{in-situ} measurements. However, for all ENLIL model runs, the simulated proton temperature is systematically too small by about an order of magnitude\footnote{The offset in the temperature values is probably caused by a scaling factor used in the model to scale the output from internal values to physical quantities. When requesting model runs on the CCMC homepage, the scaling factor cannot be modified by the user.}.

The lowest agreement between models and observations was found for the proton density and total magnetic field strength parameters with correlation coeffients $<$\,0.3.
For the proton density, the model results from MAS/ENLIL and WSA/ENLIL differ significantly in their peak values. MAS/ENLIL tends to overestimate the density peaks by up to a factor of $\approx$2, while WSA/ENLIL gives density peaks that are significantly smaller than in the observational data. For all models, the simulated total magnetic field strength is too small by at least a factor of $\approx$2. 

From the WSA/ENLIL model run with different synoptic magnetic maps (NSO, MWO, and GONG), we found that the model performance in single CRs is very sensitive to the usage of different synoptic maps. However, there is no trend as to which observatory providing the synoptic map is leading to systemically better results. Over the whole year, the differences even out and the correlation coefficients (\textit{e.g.} for the solar wind speed, cc~=~0.53) are similar for the WSA/ENLIL runs with different synoptic maps.

		\begin{figure}
			\centering
			\includegraphics[width=1\textwidth]{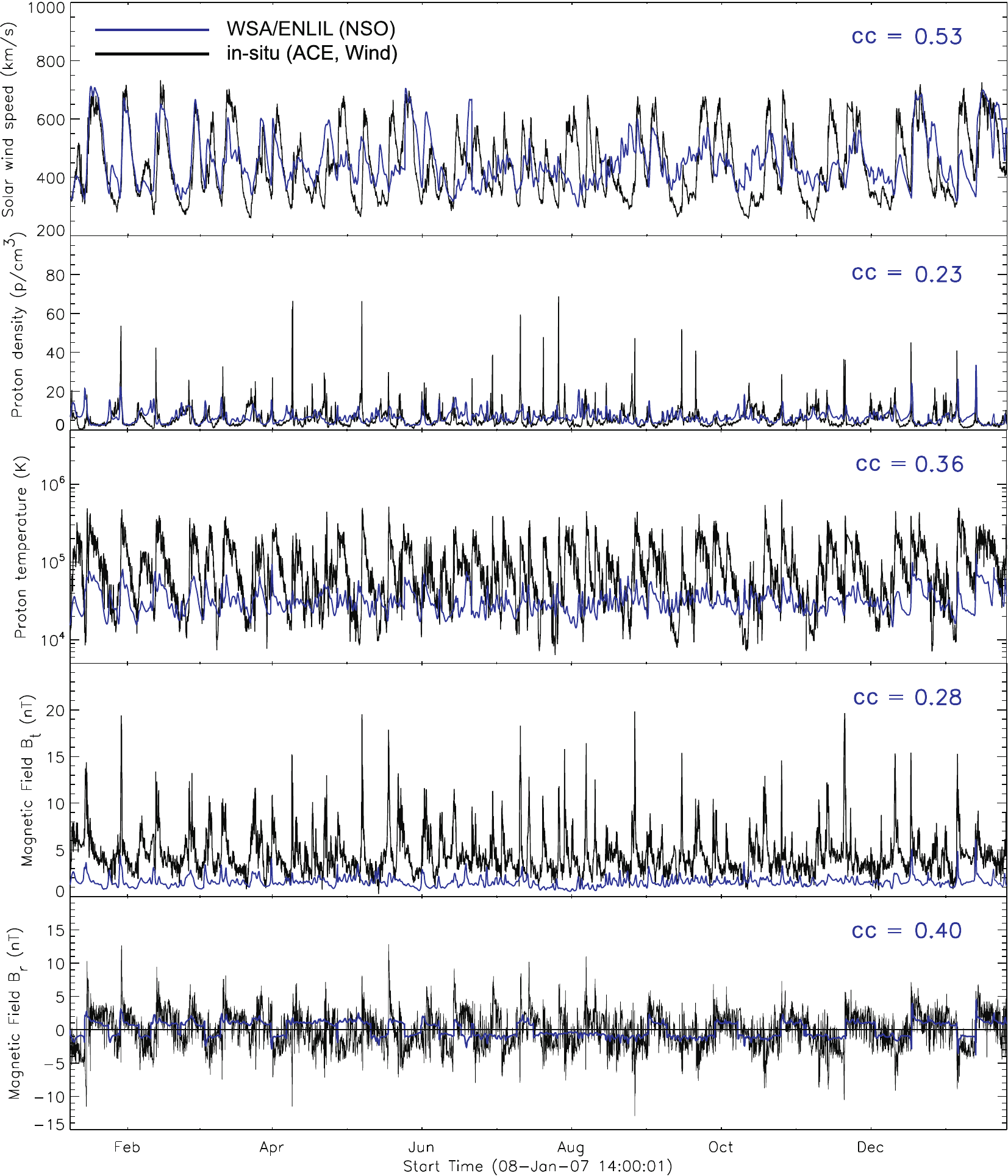}
			\caption{Modeled solar wind parameters from WSA/ENLIL (NSO, blue) and the \textit{in-situ} observations (black) for CRs~2052\,--\,2065. From top to bottom: solar wind speed, proton density, proton temperature, total magnetic field strength, and radial magnetic field strength.}
			\label{fig:nso_all_parameters}
		\end{figure}

		\begin{figure}
			\centering
			\includegraphics[width=1\textwidth]{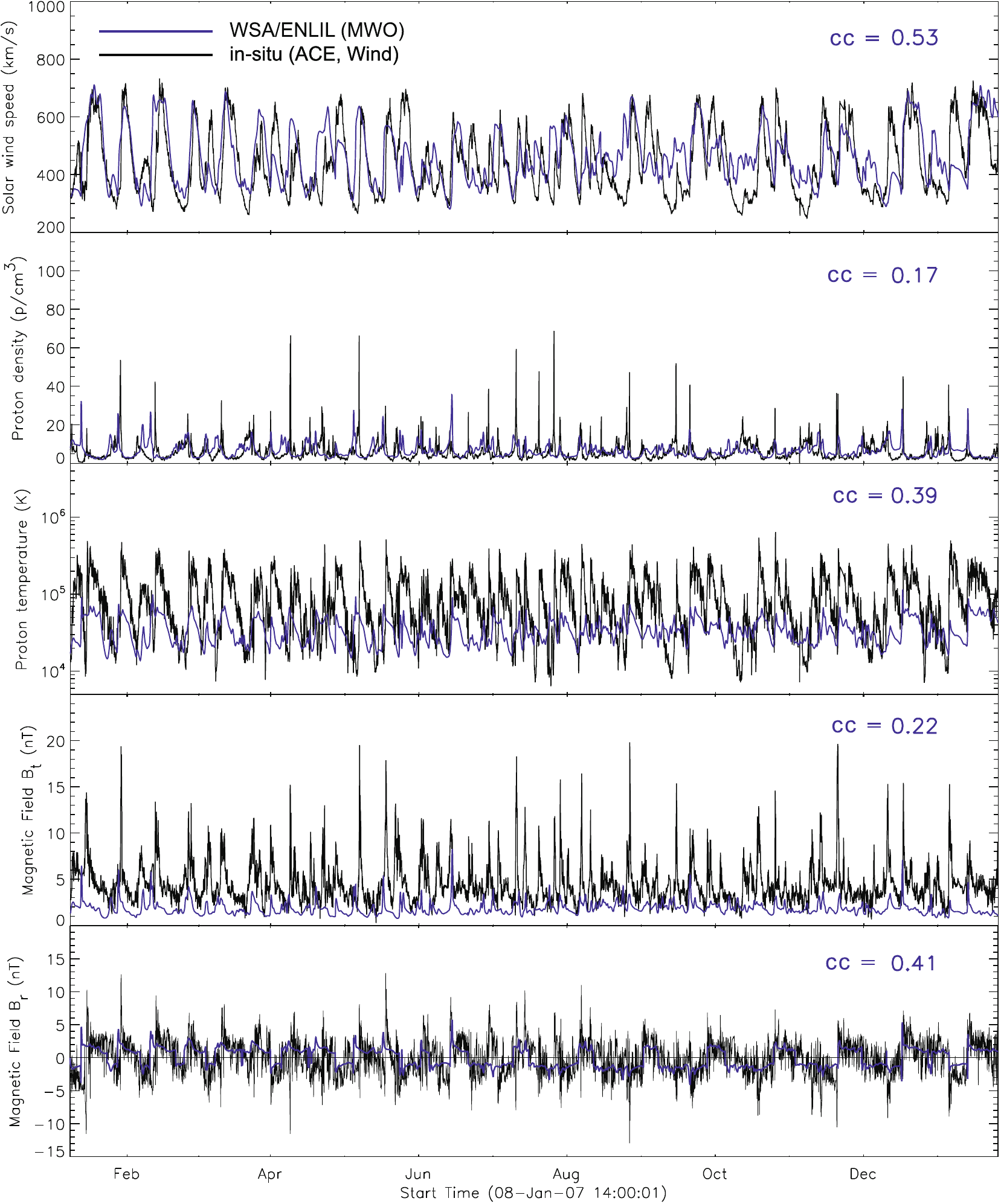}
			\caption{Modeled solar wind parameters from WSA/ENLIL (MWO, purple) and the \textit{in-situ} observations (black) for CRs~2052\,--\,2065. From top to bottom: solar wind speed, proton density, proton temperature, total magnetic field strength, and radial magnetic field strength.}
			\label{fig:mwo_all_parameters}
		\end{figure}

		\begin{figure}
			\centering
			\includegraphics[width=1\textwidth]{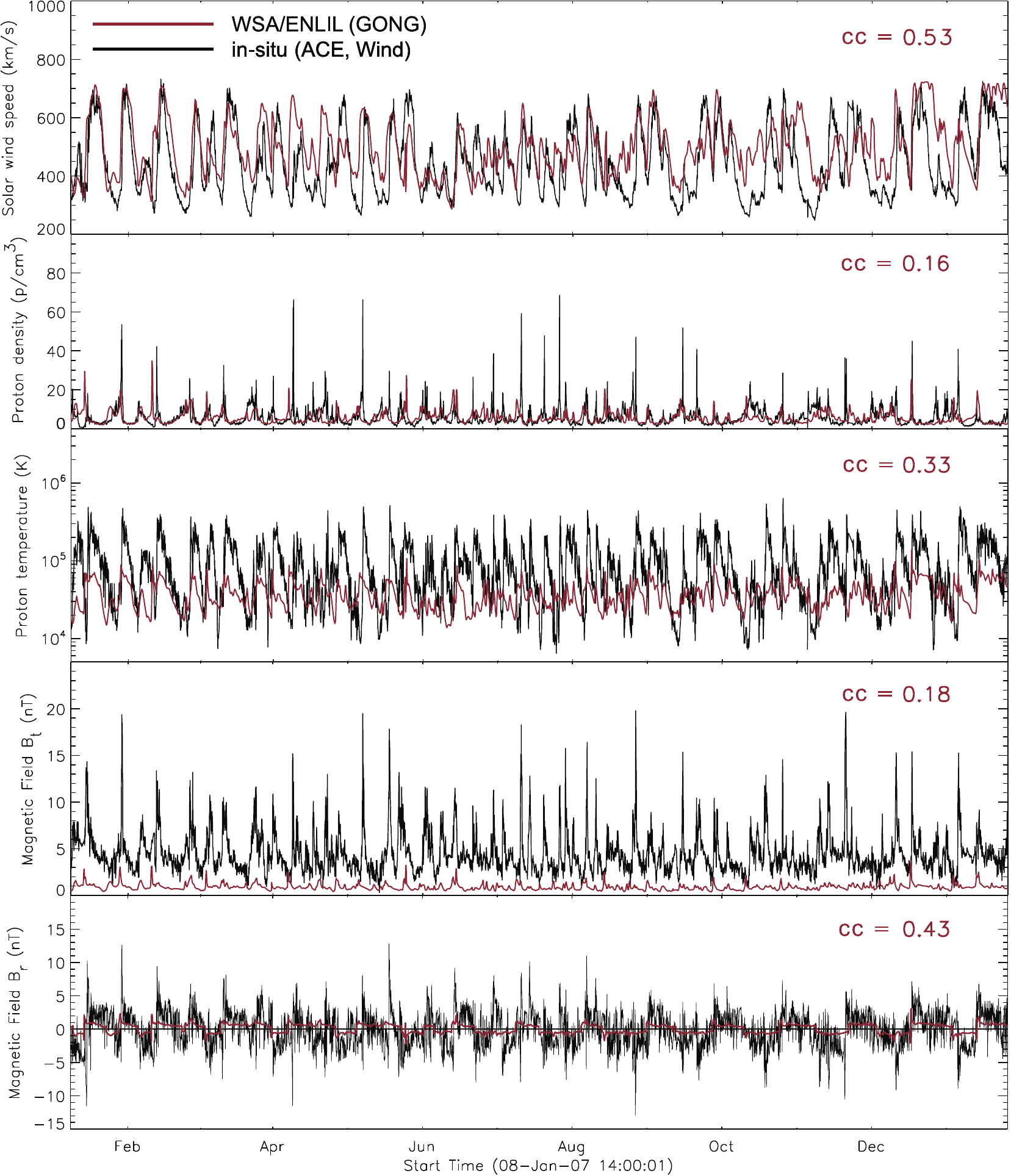}
			\caption{Modeled solar wind parameters from WSA/ENLIL (GONG, red) and the \textit{in-situ} observations (black) for CRs~2052\,--\,2065. From top to bottom: solar wind speed, proton density, proton temperature, total magnetic field strength, and radial magnetic field strength.}
			\label{fig:gong_all_parameters}
		\end{figure}

\subsection{Cross-correlation Analysis}
\label{sec:time_lag}

\begin{figure}
			\centering
			\includegraphics[width=1.\textwidth]{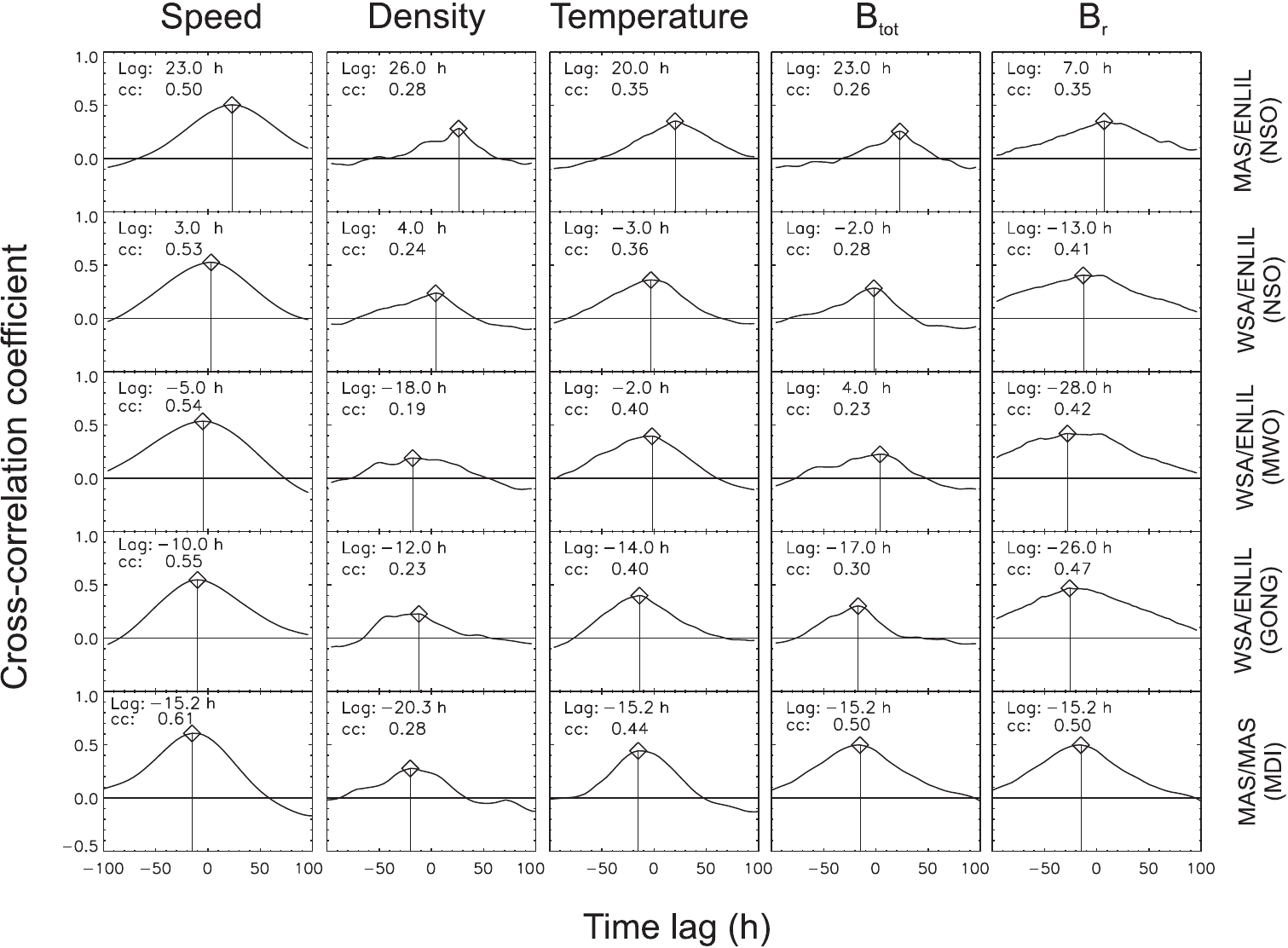}
			\caption{Results from the cross-correlation analysis for the model runs MAS/ENLIL (NSO), WSA/ENLIL (NSO), WSA/ENLIL (MWO), WSA/ENLIL (GONG), and MAS/MAS for all solar wind parameters under study.}
			\label{fig:xcorr_plots}
		\end{figure}

We were especially interested in how well the models predict the arrival time of solar wind structures at Earth. In Figure~\ref{fig:xcorr_plots}, we present the results from the cross-correlation analysis over the whole time period of CRs~2052\,--\,2065 (year 2007) for all of the model runs and all parameters. Each cross-correlation plot gives the characteristic time lag between modeled and measured arrival times of solar wind structures at Earth, as well as the maximum cross-correlation coefficient (correlation coefficient at time lag of best agreement). The highest cross-correlation coeffients (cc~$>$~0.5) are obtained for the parameter solar wind speed. The correlation coefficients, time lags, and cross-correlation coefficients for the solar wind speed parameter are summarized in Table \ref{tab:cc}.

\begin{figure}
			\centering
			\includegraphics[width=1\textwidth]{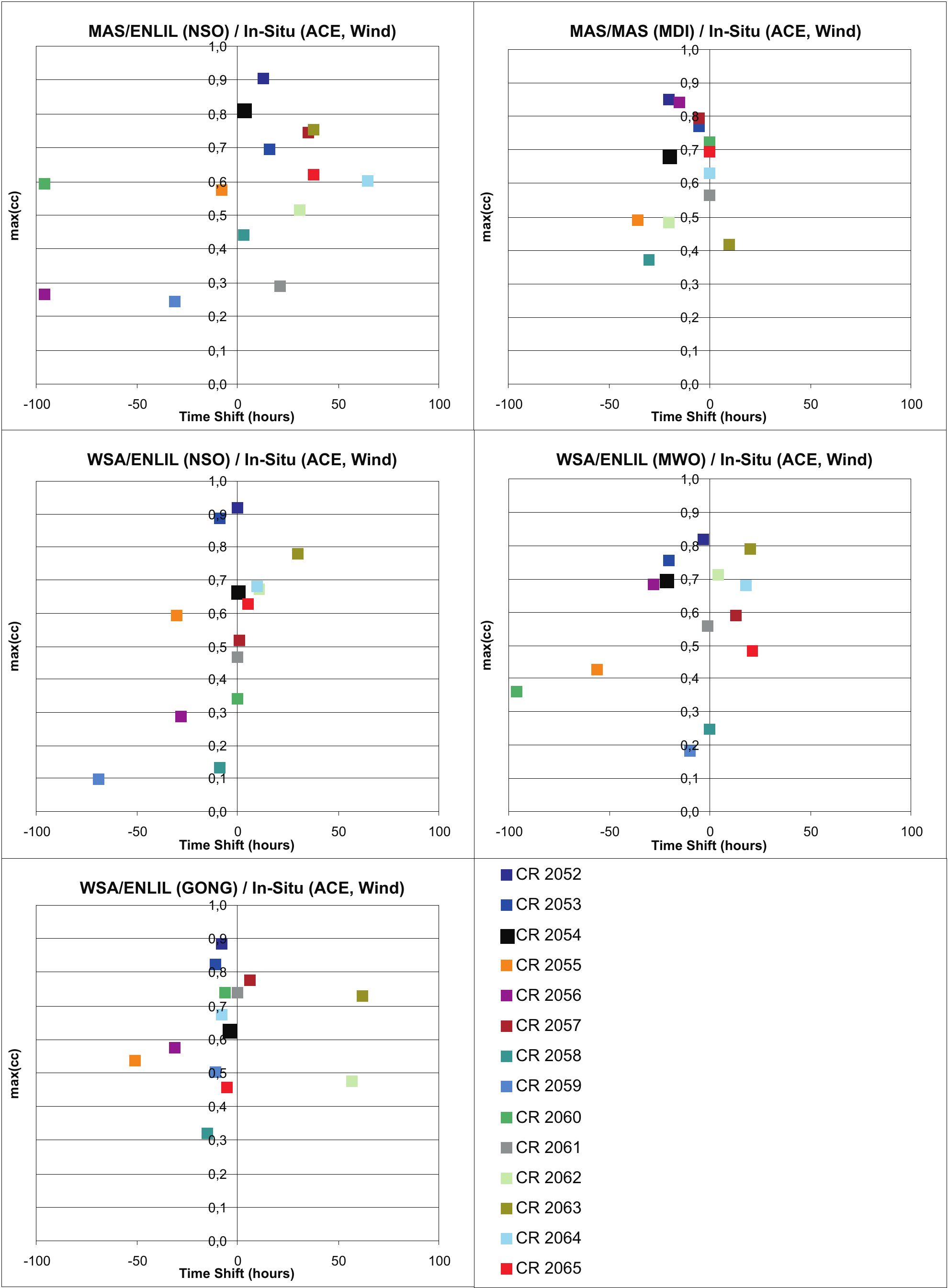}
			\caption{Results of the cross-correlation analysis for the modeled solar wind speed versus the \textit{in-situ} data. We plot for five different models the maximum cross-correlation coefficients obtained for each CR against the time lags of the highest cross-correlation.}
			\label{fig:excel_2007}
		\end{figure}

For the solar wind speed parameter, the results of the cross-correlation analysis obtained for single CRs are presented in Figure~\ref{fig:excel_2007} and Table~\ref{tab:shift_speed}. The summary plots show the time lags and cross-correlation coefficients derived from all the single CRs during 2007. In Table~\ref{tab:shift_speed} the median and median absolute deviation (MAD) as well as the mean value and standard deviation of the results are listed. Figure~\ref{fig:excel_2007} shows a tendency for the MAS/ENLIL model to be shifted to the positive for single CRs, \textit{i.e.} the model predicts the arrival of characteristic solar wind structures systematically too late, and to include several outliers with time lags $>$\,two~days. For MAS/ENLIL we obtained a mean time lag of 14.5~$\pm$\,21.5~hours and a mean cross-correlation coefficient cc~=~0.60~$\pm$~0.15. This means that on average MAS/ENLIL predicts the arrival of solar wind structures with a typical uncertainty of 21.5~hours at a systematic trend of $-$\,14.5~hours.
For WSA/ENLIL (NSO) we found no systematical time shift (0.0~$\pm$9.0~hours, cc~=~0.61~$\pm$~0.16). The WSA/ENLIL model runs with synoptic maps from MWO give a time lag of $-$2.0~$\pm$\,19.0~hours (cc~=~0.64~$\pm$~0.14), and the runs with synoptic maps from GONG give a time lag of $-$7.0$\,\pm$\,6.0~hours (cc~=~0.65~$\pm$~0.12). The results from the MAS/MAS model show a time lag of $-$5.0~$\pm$~8.0~hours and the highest cross-correlation coefficient cc~=~0.69~$\pm$~0.12.

		\clearpage

\begin{table}[t]
		\begin{tabular}{ l | c  r  c }
				\hline
			 	Model							& Correlation					&Time Lag~					&	Cross-Correlation \\
			 										& coefficient					& [hours]~~~						&	coefficient \\
				\hline
				MAS/ENLIL (NSO)		&	0.43	& 23.0~~~ 		&	0.50 \\
				WSA/ENLIL (NSO)		& 0.53	&	3.0~~~			&	0.53 \\ 
				WSA/ENLIL (MWO)		& 0.53	& $-$5.0~~~ 	& 0.54 \\
				WSA/ENLIL (GONG)	& 0.53	& $-$10.0~~~ & 0.55 \\
				MAS/MAS (MDI)			& 0.57 	& $-$15.2~~~	& 0.61 \\
				\hline
		\end{tabular}
	\caption{Correlation coefficients, time lags, and cross-correlation coefficients for the parameter solar wind speed derived from cross-correlation analysis between modeled and observed data over the whole of the year 2007 plotted in Figure~\ref{fig:xcorr_plots}.}
	\label{tab:cc}
\end{table}


		\clearpage

\begin{table}[t]
		\begin{tabular}{ l | r  c  r  c }
				\hline
				& \multicolumn {2}{|c|}{\textbf{Median} } & \multicolumn {2}{|c}{\textbf{Mean} } \\
			 	\textbf{Model}		& Time Lag~~					&	Cross-Corr. & Time Lag~~			& Cross-Corr. \\
			 										& 	[hours]~~~				&	coefficient				&   [hours]~~~		& coefficient \\
				\hline
				2007 (CR~2052\,--\,2065)&								&	\\
				MAS/ENLIL (NSO)		& 14.5 $\pm$\,21.5		& 0.60 $\pm$\,0.15 	& 2.3 $\pm$\,47.7 			& 0.57 $\pm$\,0.21 \\
				WSA/ENLIL (NSO)		& 0.0 $\pm$\,~9.0			& 0.61 $\pm$\,0.16  & $-$6.3 $\pm$\,23.5 		& 0.55 $\pm$\,0.26 \\
				WSA/ENLIL (MWO)		& $-$2.0 $\pm$\,19.0	& 0.64 $\pm$\,0.14  & $-$11.4 $\pm$\,32.4		& 0.57 $\pm$\,0.20 \\
				WSA/ENLIL (GONG)	& $-$7.0 $\pm$\,~6.0	& 0.65 $\pm$\,0.12  & $-$1.8 $\pm$\,29.5		& 0.63 $\pm$\,0.16 \\
				MAS/MAS (MDI)			& $-$5.0 $\pm$\,~8.0	& 0.69 $\pm$\,0.12  & $-$10.1 $\pm$\,13.4		& 0.64 $\pm$\,0.16 \\
				\hline
		\end{tabular}
	\caption{Time lags and cross-correlation coefficients for the parameter solar wind speed between modeled and observed data derived from cross-correlation analysis of individual CRs plotted in Figure~\ref{fig:excel_2007}.}
	\label{tab:shift_speed}
\end{table}

\subsection{Distribution of Solar Wind Parameters}

\begin{figure}
			\centering
			\includegraphics[width=1.\textwidth]{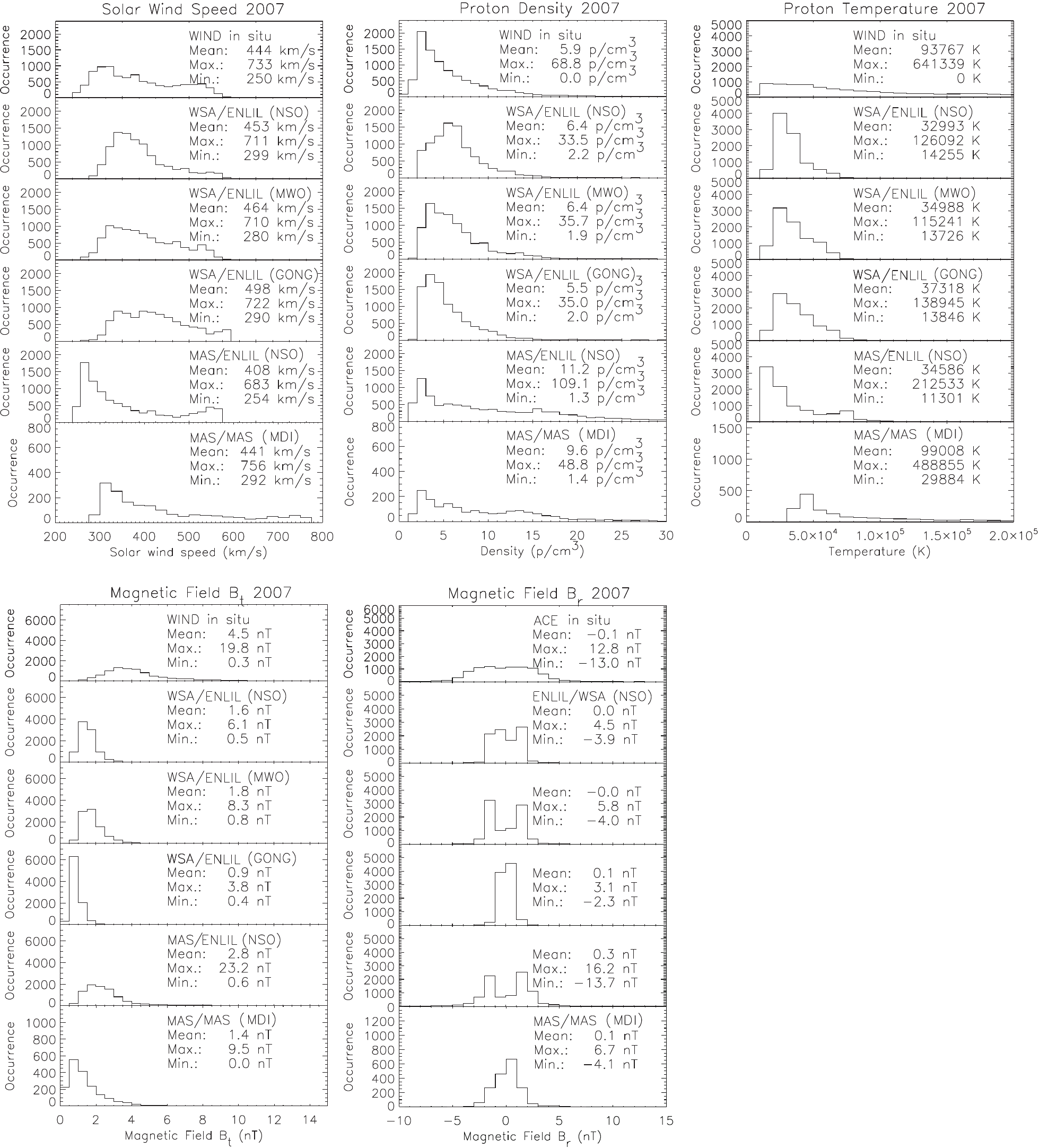}
			\caption{Distribution of solar wind proton density (binsize: 1~pcm$^{-3}$), speed (binsize: 25~kms$^{-1}$), temperature (binsize: $10^4$~K), total magnetic field (binsize: 0.5~nT), and radial magnetic field (binsize: 1~nT) for the year 2007 (CRs~2052\,--\,2065). The top panels show the histograms for the \textit{Wind} and ACE \textit{in-situ} data, the other four panels show the histograms for MAS/ENLIL(NSO), WSA/ENLIL (MWO), WSA/ENLIL (GONG), and MAS/ENLIL (NSO), and MAS/MAS (MDI) model results. The maximum, minimum, and average density are displayed on the top right of each histogram.}
			\label{fig:histogram}
		\end{figure}

In Figure~\ref{fig:histogram} we compare the distribution of the modeled solar wind parameters with the distribution gained from \textit{in-situ} measurements.
The histograms in Figure~\ref{fig:histogram} show that the models do not reproduce the characteristic distribution of solar wind parameters. MAS/ENLIL (NSO) highly overestimates the occurrence of slow speed solar wind at values around $\approx$300~kms$^{-1}$. Both MAS/ENLIL and WSA/ENLIL overestimate the occurrence of small temperatures and magnetic field strengths. For MAS/ENLIL and WSA/ENLIL the variability of the modeled solar wind parameters temperature and magnetic field strength is much smaller than what can be found from the \textit{in-situ} measurements. For WSA/ENLIL run with synoptic maps from GONG, the simulated total magnetic field strengths are extremely small and the peaks in the simulations hardly ever reach more than 2~nT, in contrast to the \textit{in-situ} data that reveal enhancements of the total magnetic field strength up to 20~nT. The MAS/MAS model shows the largest variability among the solar wind models, with the maximum and minimum speeds of the solar wind in basic agreement with the observations. However, for MAS/MAS the distribution of the solar wind speed also differs from the measured distribution: MAS/MAS underestimates the occurrence of slow speed solar wind and overestimates the occurrence of high speeds.

		\clearpage

\section{Discussion and Conclusion}

The main findings of our study are:
\begin{itemize}
	\item All solar wind models produce the best simulation results for the parameters solar wind speed with correlation coefficients between 0.4 and 0.6. 
	\item The interplanetary sector structure (as coded in the radial magnetic field strength) is well reproduced and shows correlation coefficients between 0.3 and 0.5.
	\item In general, the model results from MAS/MAS give the highest correlation coefficients (cc = 0.57 for solar wind speed and cc = 0.44 for the radial magnetic field strength).
	\item Model runs from WSA/ENLIL tested with different synoptic magnetic maps show significant differences in predicted arrival time and amplitudes of solar wind structures. However, we found no clear trend as to which synoptic map gives the best simulation results.
	\item For all models, the distributions of modeled solar wind parameters significantly differ from the measured distribution.
	\item  The predicted proton temperature at the location of Earth is too low by an order of magnitude for the MAS/ENLIL and WSA/ENLIL models. All models give too small total magnetic field strengths.
	\item The typical uncertainties in predicting the arrival of solar wind structures during quiet Sun conditions (year 2007), are in the range of 0.5\,--\,1.5~days for MAS/ENLIL and 0\,--\,1~days for WSA/ENLIL and MAS/MAS with cross-correlation coefficients in the range of 0.5 to 0.7.
\end{itemize}

The results we obtained from comparing different solar wind models to \textit{in-situ} observations are in basic agreement with previous studies by \cite{lee2009} and \cite{jian2011}.
These authors tested the model performance during times of medium to high solar activity (2003\,--\,2006). In our study, we focused on a time period that was characterised by extremely low ICME activity (2007) and should therefore yield ideal conditions for background solar wind modeling. Solar activity can worsen the accuracy of the model performance in two ways.
First, CME activity disturbs the interplanetary background solar wind structure. In this case, the effects of CMEs on the solar wind conditions are observed by the \textit{in-situ} measurements but do not appear in the predictions of the background solar wind. 
Second, the synoptic magnetic maps basically give an average of the photospheric magnetic field over the course of one CR. High solar activity, \textit{i.e.} evolving sunspots, flares, and CME eruptions, imply large changes in the solar magnetic field within one CR, which are not captured in the synoptic maps. Similar to previous studies, our comparison showed that the large-scale structure of the heliosphere is simulated well. However, even under quiet conditions, the predictions of the arrival time of solar wind structures still have typical uncertainties of the order of about one day. For space weather forecasting and propagation studies of ICMEs, simulations that are accurate on timescales $\lesssim$\,one day are required.

The magnetic field strengths are underestimated by ENLIL and MAS by a factor of approximately two.
The underestimation of the modeled magnetic field strength was analyzed in detail by \cite{stevens2012}. These authors showed that the underestimation of the magnetic flux is caused by too low density and temperature values at the base of the coronal model and by the restrictions coming from the computational-grid resolution of the model runs. For modeling near the ecliptic plane, where the relatively thin interplanetary current sheet is located, the model resolution is of high importance for more accurate simulations. \cite{stevens2012} showed that by varying the base temperature and density and increasing the model resolution, more realistic magnetic field strengths can be obtained.

An important outcome is that the simulation results differ widely when using different combinations of synoptic map, coronal and heliospheric model. However, we found no trend as to which model combination gives systematically better simulation results than the other. A certain coronal/heliospheric model combination may have a low agreement with the observations during one CR, but reveal a good model performance for the following CR. Similarly, although the source of the synoptic map largely affects the model results, there is no trend as to which source leads to the best simulation results. In accordance with the findings of \cite{riley2012} we find that the model performance is very sensitive to the input synoptic maps. Therefore, a promising way to improve model results would be to do a detailed study on how the usage of synoptic maps from different observatories and the way they are processed influence the model performance.

%

%
 \begin{acks}

We acknowledge the use of \textit{Wind} data provided by the solar wind experiment teams at GSFC. We thank the ACE SWEPAM and MAG instrument team and the ACE Science Center for providing the ACE data.
Simulation results for the ENLIL model have been provided by the Community Coordinated Modeling Center at Goddard Space Flight Center through their public Runs on Request system (\url{ccmc.gsfc.nasa.gov)}.
The model runs for the MAS model were carried out by Predictive Science Inc.\ (\url{www.predsci.com/portal/home.php}).
The research leading to these results has received funding from the European Commission FP7 Project COMESEP (project n$^{\circ}$ 263252, \url{comesep.aeronomy.be}) and
from the European Commission's Seventh Framework Programme (FP7/2007-2013) under the grant agreement eHeroes (project n$^{\circ}$ 284461, \url{www.eheroes.eu}). M.\,Temmer acknowledges the Austrian Science Fund (FWF): FWF V195-N16.

 \end{acks}

%
%
 \bibliographystyle{spr-mp-sola}

\def\newblock{\hskip .11em plus .33em minus .07em}

%
%
%
%

\end{article} 
\end{document}